# Phase Shift Analysis of the p$^{14}$C Scattering at the Energy of the $^2S_{1/2}$ Resonance

S. B. Dubovichenko[1,2,*], A. V. Dzhazairov-Kakhramanov[1,2,‡]

[1]*V. G. Fessenkov Astrophysical Institute "NCSRT" NSA RK, 050020, Observatory 23, Kamenskoe plato, Almaty, Kazakhstan*
[2]*Institute of Nuclear Physics CAE MINT RK, 050032, str. Ibragimova 1, Almaty, Kazakhstan*
[*]dubovichenko@mail.ru
[‡]albert-j@yandex.ru

The phase shift analysis for determination of the $^2S_{1/2}$ resonance at 1.5 MeV was carried out on the basis of the known experimental measurements of the excitation functions of the p$^{14}$C elastic scattering at four angles from 90° to 165° and more than 100 energy values in the range from 600–800 to 2200–2400 keV. The obtained shape of the resonance $^2S_{1/2}$ phase shift coincides with the characteristics of the corresponding level of $^{15}$N in whole, and the suggested potential acceptably describe this scattering phase shift.

## 1. Introduction

Along with the interactions of the bound states (BSs) of clusters it is necessary to know the potentials of the p$^{14}$C elastic scattering including the potential of the resonance $^2S_{1/2}$ wave at 1509(4) keV with the big width 405(6) keV and $J^\pi = 1/2^+$ (for $^{14}$C it is known that $J^\pi = 0^+$) [5] for carrying out the calculations for some reactions of the radiative proton capture on $^{14}$C in the frame of the modified potential cluster model (MPCM) at astrophysical energies [1-4]. The noticed above reaction of the radiative proton capture on $^{14}$C plays, evidently, a certain role at the nucleosynthesis of elements in the Universe at different stages of its formation. For example in work [6], it was supposed that baryon number fluctuations in the early Universe lead to the formation of high-density proton-rich and low-density neutron-rich regions [7]. This might be the result of the nucleosynthesis of elements with mass $A \geq 12$ in the neutron-rich regions of the early Universe [8,9]. The special interest is the nucleus $^{14}$C [10], which is produced by successive neutron capture on $^{12}$C [11]

$$\ldots\ ^{12}C(n,\gamma)^{13}C(n,\gamma)^{14}C\ \ldots\ . \qquad (1)$$

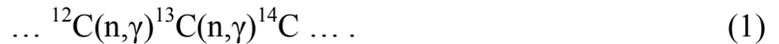

The nucleus $^{14}$C has a half-life about 5700 years and is stable on the time scale of the Big-Bang nucleosynthesis. Therefore the synthesis of elements with mass $\geq 14$ depends on the rate of the neutron, alpha, and proton capture reactions on $^{14}$C [7] and the consideration of the similar reactions seems to be interesting with a point of view of nucleosynthesis of different chemical elements in our Universe. [12].

Passing to the direct description of the results for our phase shift analysis of the p$^{14}$C elastic scattering let us note that earlier we already have performed the phase shift analysis in systems p$^6$Li [13], n$^{12}$C [14], p$^{12}$C [15], $^4$He$^4$He [16], $^4$He$^{12}$C [17], p$^{13}$C [18], and n$^{16}$O [19], meanwhile, essentially at astrophysical energies [20].

## 2. Methods of calculation

The nuclear scattering phases obtained on the basis of the experimental differential cross sections allow one to extract certain information about the structure of resonance states of light atomic nuclei [21]. In this case, the processes of the elastic scattering of particles with total spin of 1/2 on the nucleus with zero spin take place in nuclear systems like $N^4He$, $^3H^4He$, $N^{12}C$, $N^{16}O$, $N^{14}C$ etc. The cross section of the elastic scattering of such particles is presented in the simple form [22]

$$\frac{d\sigma(\theta)}{d\Omega} = |A(\theta)|^2 + |B(\theta)|^2, \qquad (2)$$

where

$$A(\theta) = f_c(\theta) + \frac{1}{2ik}\sum_{L=0}^{\infty}\{(L+1)S_L^+ + LS_L^- - (2L+1)\}\exp(2i\sigma_L)P_L(\cos\theta) ,$$

$$B(\theta) = \frac{1}{2ik}\sum_{L=0}^{\infty}(S_L^+ - S_L^-)\exp(2i\sigma_L)P_L^1(\cos\theta),$$

$$f_c(\theta) = -\left(\frac{\eta}{2k\sin^2(\theta/2)}\right)\exp\{i\eta\ln[\sin^{-2}(\theta/2)] + 2i\sigma_0\} . \qquad (3)$$

Here $S_L^\pm = \eta_L^\pm \exp(2i\delta_L^\pm)$ – scattering matrix, $\delta_L^\pm$ – required scattering phase shifts, $\eta_L^\pm$ – inelasticity parameters, and signs "±" correspond to the total moment of system $J = L \pm 1/2$, $k$ – wave number of the relative motion of particles $k^2 = 2\mu E/\hbar^2$, $\mu$ – reduced mass, $E$ – the energy of interacting particles in the center-of-mass system, $\eta$ – Coulomb parameter.

The multivariate variational problem of finding these parameters at the specified range of values appears when the experimental cross sections of scattering of nuclear particles and the mathematical expressions, which describe these cross sections with certain parameters $\delta_L^J$ – nuclear scattering phase shifts, are known. Using the experimental data of differential cross-sections of elastic scattering, it is possible to find a set of phase shifts $\delta_L^J$, which can reproduce the behavior of these cross-sections with certain accuracy. Quality of description of experimental data on the basis of a certain theoretical function or functional of several variables (2,3) can be estimated by the $\chi^2$ method, which is written as

$$\chi^2 = \frac{1}{N}\sum_{i=1}^{N}\left[\frac{\sigma_i^t(\theta) - \sigma_i^e(\theta)}{\Delta\sigma_i^e(\theta)}\right]^2 = \frac{1}{N}\sum_{i=1}^{N}\chi_i^2, \qquad (4)$$

where $\sigma^e$ and $\sigma^t$ are experimental and theoretical, i.e., calculated for some defined values of the scattering phase shifts cross-sections of the elastic scattering of nuclear particles for $i$-angle of scattering, $\Delta\sigma^e$ – the error of experimental cross-sections at these angles, $N$ – the number of measurements. The details of the using by us searching method of scattering phase shifts were given in [22] and in our works [16,20].

The next values of particle masses are used in the given calculations: $m_p = 1.00727646577$ amu [23], and $m(^{14}C) = 14.003242$ amu [24], and constant $\hbar^2/m_0$ is equal to 41.4686 MeV fm$^2$.

## 3. Phase shifts and potentials of the elastic scattering

The excitation functions from the work [7], measured at 90°, 125°, 141° and 165° in the energy range from 0.6 to 2.3 MeV (l.s.), are shown in Figs. 1a,b,c,d by dots. These data are used by us further for carrying out of the phase shift analysis and extracting of the resonance form of the $^2S_{1/2}$ scattering phase at 1.5 MeV. The results of the present analysis are shown in Figs. 2a,b,c,d by dots, and the solid lines in Fig. 1 show cross sections calculated with the obtained scattering phase shifts. About 120 first points, cited in work [7], in the stated above energy range were used in this analysis. In addition, it was obtained that for description of the cross sections in the excitation functions, at least at the energies up to 2.2–2.3 MeV, there is no need to take into account $^2P$ or $^2D$ scattering waves, i.e., their values simply can be equalized to zero. Since only one point in the cross sections of excitation functions is considered for each energy and angle, therefore the value of $\chi^2$ at all energies and angles usually lays at the level $10^{-2}$–$10^{-10}$ and taking into account another partial scattering phase shifts already does not lead to its decrease.

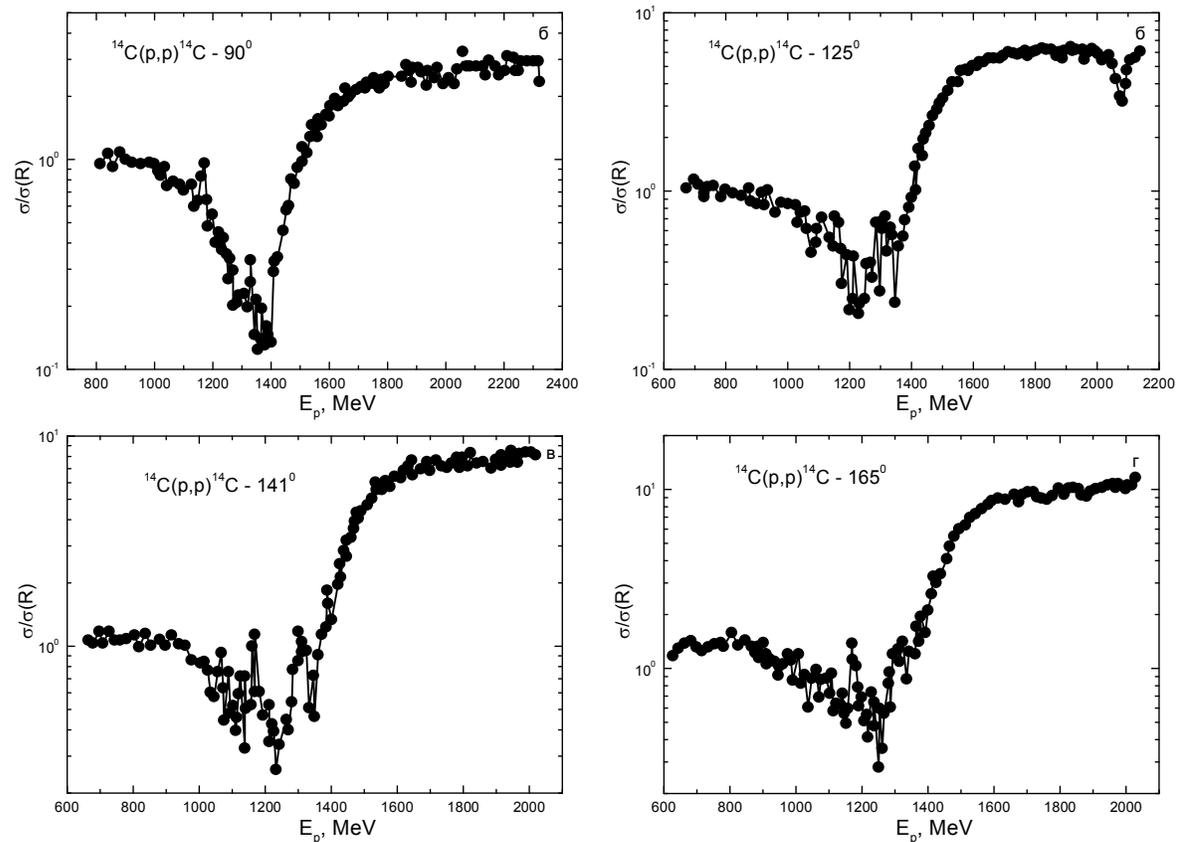

Figs. 1a,b,c,d. The excitation functions in the elastic p$^{14}$C scattering in the range of the $^2S_{1/2}$ resonance [7]. The solid line – their approximation on the basis of the obtained scattering phase shifts.

The resonance energy, as it is seen in Fig. 2, obtained from the excitation function at 90° is at the interval of 1535–1562 keV for which the phase shift value lays within limits of 87°–93° with the value of 90° at 1554 keV. The resonance energy obtained from the excitation function at 125° is at the interval of 1551–1575 keV for which the phase shift value lays within limits of 84°–93°. The resonance energy obtained from the excitation function at 141° is at the interval of 1534–1611 keV for which the phase shift value lays within limits of 84°–90° with the value of 90° at 1534, 1564 and 1611 keV. The resonance energy obtained from the excitation function at 165° is at the interval of 1544–1563 keV for which the phase shift value lays within limits of 87°–91°. For so noticeable scatter of values, it can be said only that the resonance value lays within limits of 1534–1611 that is, in general, agree with data [5], where the resonance energy value of 1509 keV is given.

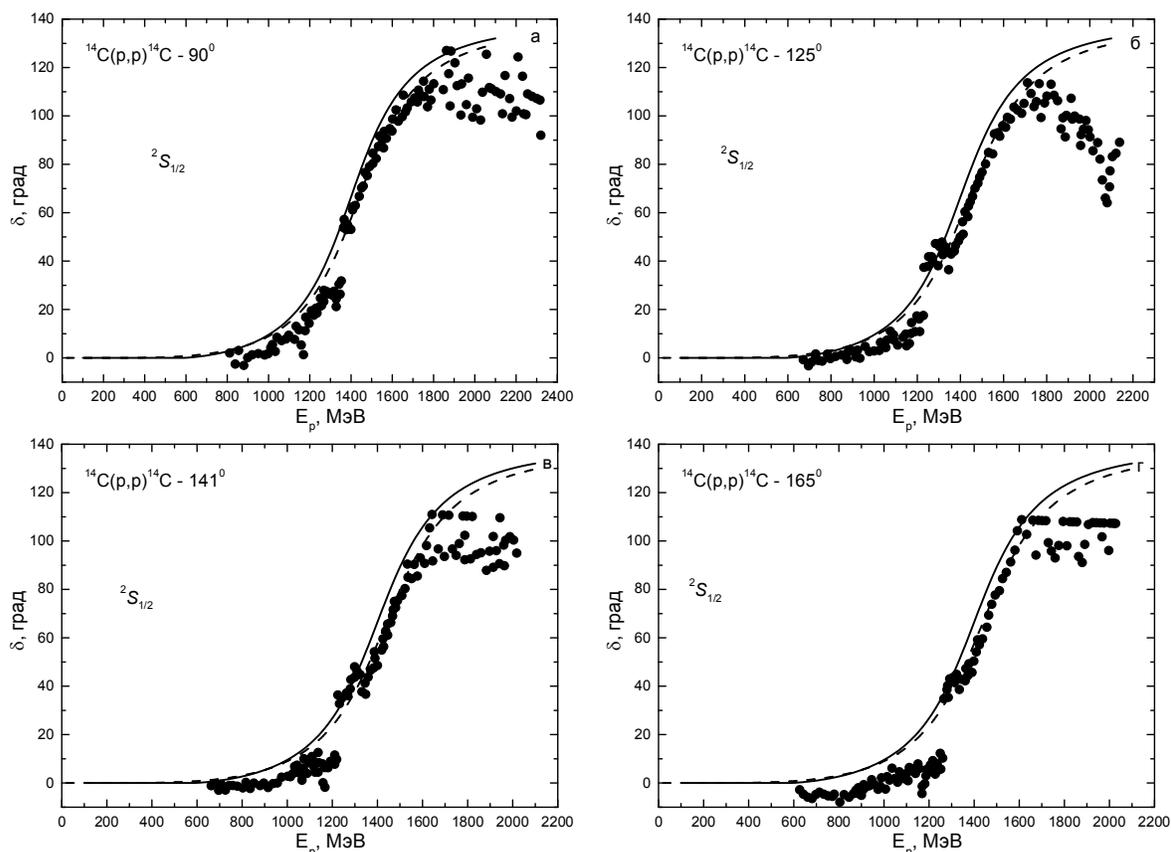

Figs. 2a,b,c,d. The $^2S_{1/2}$ elastic phase shift of the p$^{14}$C scattering at low energies, obtained on the basis of the excitation functions, shown in Figs. 1a,b,c,d. Points – results of our phase shift analysis, carried out on the basis of data from [50], lines – calculation of the phase shift with the potentials given in the text.

Let us note that work [7] mentions that the detailed analysis of the resonances, including 1.5 MeV, was not carried out, because it was done earlier in works [25] on the basis of the proton capture reaction on $^{14}$C. Here, as one can see from the results of the phase shift analysis, the resonance energy at 1.5 MeV slightly overestimated. However, as it was seen in Fig. 1, the data spread on cross sections in excitation functions is too large for clear conclusion about the energy of resonance. Apparently, the additional and more modern measurements of the elastic scattering cross sections are required, in order to on the basis of these data to perform more unambiguous conclusion about the resonance energy at 1.5 MeV.

Going to the construction of the potentials for the p$^{14}$C elastic scattering, let us consider, at first, the classification of orbital states on Young tableaux thinking that the tableau {4442} correspond to the ground bound state (GS) of $^{14}$C [3,4]. Possible orbital Young tableaux in the system of $N = n_1 + n_2$ particles can be determined as a direct external product of the orbital tableaux of each subsystems, in the case of p$^{14}$C system it gives {1} × {4442} → {5442} + {4443} [3,4]. The firs from the obtained tableaux is compatible with the orbital moment $L = 0$ and is forbidden, because it should not be five nucleons in the $s$-shell, and the second tableaux is allowed and compatible with the orbital moment equals unit [3,4]. Thereby, the forbidden bound state (FS) there is in the potential of the $^2S_{1/2}$ wave, and the $^2P$ waves have only allowed states (ASs). The ground state of $^{15}$N in the p$^{14}$C channel, which is at the energy of -10.2074 MeV [5], related to the $^2P_{1/2}$ wave and also does not contain FSs. However, because we haven't total tables of the product of Young tableaux for systems with the number of particles more than eight [26], which are used by us for the similar calculations earlier [1-4,20,27-29], so the result obtained above should be considered only the qualitative estimation of possible orbital symmetries for the BS of $^{15}$N in the p$^{14}$C channel.

For description of the obtained $^2S_{1/2}$ scattering phase in the phase shift analysis it is possible to use simple Gaussian potential of the form:

$$V(r) = -V_0 \exp(-\alpha r^2) \qquad (5)$$

with FSs and parameters

$$V_0 = 5037.0 \text{ MeV}, \quad \alpha = 12.0 \text{ fm}^{-2}, \qquad (6)$$

which lead to the scattering phase shifts with the resonance at 1500 keV (l.s.) and with the width of 530 keV (c.m.) that is in a good agreement with the available experimental data [5]. The parameters of this potential were matched to reproduce in general the resonance data exactly from [5], which were obtained in works [25]. The phase shift of this potential is shown in Figs. 2a,b,c,d by the solid lines and at the resonance energy reaches the value of 90(1)°. The energy behavior of the scattering phase shift of this potential correctly describes obtained in the phase shift analysis scattering phases in whole, taking into account the shift of the resonance energy approximately at 30–50 keV relative to results [5]. The calculated phase shift line for this potential is parallel to points, obtained in our phase shift analysis, for all scattering angles.

For more accurate description of the obtained in the phase shift analysis data the next potential is needed

$$V_0 = 5035.5 \text{ MeV}, \quad \alpha = 12.0 \text{ fm}^{-2}. \qquad (7)$$

It leads to the resonance energy of 1550 keV, its width of 575 keV, and the calculation results of the $^2S_{1/2}$ phase shift are shown in all Figs. 2 by the dashed line. As one can see this line appreciably better reproduce results of the carried out here phase shift analysis.

It should be noted over again that the potential is constructed completely unambiguously, if the number of FSs is given (in this case, it is equal to unit), according to the known energy of the resonance level in spectra of any nucleus [5] and its width. In other words, it is not possible to find another combination of the parameters $V$ and α, which could be possible to describe the resonance energy of level and its width correctly. The depth of such potential unambiguously determines the resonance location, i.e., resonance energy of the level, and its width α specifies the certain width of this resonance state, which have to correspond to experimental observable values [5].

## 4. Conclusion

Thereby, the resonance $^2S_{1/2}$ phase shift of the p$^{14}$C elastic scattering at energies from 0.6 MeV to 2.3 MeV was found as a result of the carried out phase shift analysis of the experimental differential cross sections in excitation functions [7]. The resonance energy of the phase shift is in a quite agreement with the level spectrum of $^{15}$N [5] in the p$^{14}$C channel. The results of the carried out phase shift analysis, i.e., phase shift of the elastic p$^{14}$C scattering and the data of resonances of $^{15}$N [5], allow one to parametrize intercluster interaction potentials for scattering processes in the resonance $^2S_{1/2}$ wave. These potentials, by-turn, can be used further for carrying out certain calculations for different astrophysical problems, partially considered, for example, in [2,3,20].

It should be noted that it is not enough data of work [7] for carrying out of the accurate phase shift analysis and the another additional data on differential cross sections, for example, angular distributions in the resonance region are needed. Such data can allow one to determine the location of the $^2S_{1/2}$ resonance in the region of 1.5 MeV and more accurately determine its width exactly on the basis of the elastic scattering phase shifts.